\documentclass[sigconf,nonacm]{acmart}

\usepackage[most]{tcolorbox}
\usepackage{cleveref}
\usepackage{booktabs}
\usepackage{makecell}
\usepackage{xspace}
\usepackage{subcaption}

\newcommand{\name}{document testing\xspace}

\AtBeginDocument{%
  }

\begin{document}

\title[Identifying Inaccurate Descriptions in LLM-generated Code Comments via Test Execution]{Identifying Inaccurate Descriptions in \\ LLM-generated Code Comments via Test Execution}

\author{Sungmin Kang}
\email{sungmin.kang@kaist.ac.kr}
\affiliation{%
  \institution{KAIST}
  \city{Daejeon}
  \country{South Korea}
}

\author{Louis Milliken}
\email{lmilliken@kaist.ac.kr}
\affiliation{%
  \institution{KAIST}
  \city{Daejeon}
  \country{South Korea}
}

\author{Shin Yoo}
\email{shin.yoo@kaist.ac.kr}
\affiliation{%
  \institution{KAIST}
  \city{Daejeon}
  \country{South Korea}
}

\renewcommand{\shortauthors}{Kang et al.}

\begin{abstract}
  Software comments are critical for human understanding of software, and as such
  many comment generation techniques have been proposed. However, we find that
  a systematic evaluation of the factual accuracy of generated comments is rare; only
  subjective accuracy labels have been given. Evaluating comments generated by
  three Large Language Models (LLMs), we find that even for the best-performing LLM, roughly
  a fifth of its comments contained demonstrably inaccurate statements. While
  it seems code-comment consistency detection techniques should be able to
  detect inaccurate comments, we perform experiments demonstrating they have 
  no statistically significant relationship  with comment accuracy, underscoring the substantial
  difficulty of this problem. To tackle this, we propose the concept of
  \name, in which a document is verified by using an LLM to generate tests
  based on the document, running those tests, and observing whether they pass or fail. 
  Furthermore, we implement our concept to verify Java comments.
  Experiments demonstrate that our approach has a robust statistical
  relationship with comment accuracy, making headway into a problem where prior
  techniques failed. Qualitative evaluation also reveals the promise of our approach
  in gaining developer trust, while highlighting the limitations of our current
  implementation.
\end{abstract}



\received{20 February 2007}
\received[revised]{12 March 2009}
\received[accepted]{5 June 2009}

\maketitle

\section{Introduction}
\label{sec:introduction}

Software documentation is an important artifact for developers as they navigate complicated repositories of code. Documentation, and code comments in particular, are instrumental for code comprehension~\cite{woodfield1981effect}. To aid developer comprehension, many automatic comment generation techniques have been proposed~\cite{mu2023dome,geng2024llmdoc,hu2018deep}. 

A critical aspect of code comments is their factual accuracy~\cite{aghajani2020docsurvey} - automatically generated comments in particular should seek to faithfully describe the behavior of the code, as inaccurate descriptions can lead to bug introduction~\cite{tan2007icomment}. It is thus surprising that most comment generation literature does not evaluate this aspect. Instead, automated documentation generation techniques are generally evaluated using similarity-based metrics such as BLEU, or more recently SentenceBERT~\cite{haque2022semantic}. However, similarity does not ensure accuracy - as with code, the modification of a single critical token in a comment has the potential to mislead the developer. Given the absence of literature that evaluates the factual accuracy of automatically generated documents, despite the importance of doing so, we evaluated the accuracy of automatically generated comments from three large language models (LLMs), which are known to be the state-of-the-art in comment generation~\cite{geng2024llmdoc}. 

Manual inspection of 540 automatically generated comments revealed that even the best-performing LLM in our study, GPT-4, generated statements that incorrectly describe the program's intent or behavior in about one fifth of its generated comments. While this is expected to a certain extent, as LLMs are known to make up facts or hallucinate~\cite{shuster2021hallucination}, 
the statistic suggests it would be difficult to use LLMs (or their generally less effective machine learning counterparts) at scale to generate comments; it is difficult to imagine a repository or file having a significant proportion of inaccurate comments and still being acceptable. This evaluation additionally provides a level of clarity that previous comment generation techniques did not - while it is unclear what a BLEU score of 50 means in practice, it is easy to understand results such as `18\% of comments generated by GPT-4 contained inaccurate content'.

It is natural to wonder whether such inaccurate comments can simply be filtered out using existing work on code-comment consistency detection techniques~\cite{panthaplackel2021deep}, which are designed to solve this particular problem. However, our evaluation of existing techniques reveals that filtering out factually inaccurate comments generated by LLMs is in fact a difficult problem. Evaluating nine techniques, specifically four state-of-the-art code-comment consistency detection techniques~\cite{Dau2024DocChecker,panthaplackel2021deep,li2024mutation}, four similarity measures~\cite{papineni2002bleu, wang2021codet5, feng2020codebert, haque2022semantic}, and one LLM response inaccuracy detector~\cite{tanzil2024cid}, we find that none of them showed a statistically significant difference in output between accurate and inaccurate documents.

What mistakes could LLMs be making when generating comments, such that they are not being detected by code-comment consistency techniques? To understand this in greater depth, we categorize the mistakes that LLMs make to construct a taxonomy of common mistake patterns. We observe four main types of errors from LLM-generated comments for Java methods, ranging from hallucinating the intent of a code function to inaccurately describing the method behavior. While some errors are difficult to automatically detect and redeem, such as inaccurately describing the intent of a function, others directly relate to the behavior of code, and thus can be tested against the existing implementation.

For these inaccurate comments that relate to the program semantics, we propose the concept of \emph{\name}: specifically, that a simple way of testing comment veracity may be to generate new tests using LLMs based on these comments, and observing the pass/fail results of the tests - simply put, to test the generated comments. The intuition is that documents that accurately describe the program behavior will lead to more passing tests, and vise versa; from this, we may be able to infer which documents are accurate based on the test results. We describe our implementation of the \name idea, and our pipeline to minimize the amount of noise that is involved in this test generation process.

Experimental results show that, unlike the baselines which could not distinguish between accurate and inaccurate documents, there was a strong statistical difference between test pass rate between accurate comments and those that inaccurately described code behavior. In turn, this difference could be used to distinguish accurate documents. Additionally, a qualitative evaluation of our results shows that \name can identify specifically which property of a comment is wrong and demonstrate its findings through tests, but it still has room for improvement.

In summary, our contributions are as follows:

\begin{itemize}
    \item We generate 540 method-level comments from Java source code using LLMs, and manually label which comments are demonstrably inaccurate when referencing the code, revealing that even the best performing LLM makes errors in roughly a fifth of its generated comments;
    \item We evaluate nine approaches based on existing work and find that none have a meaningful statistical relationship with comment correctness, highlighting the difficulty of detecting mistakes in automatically generated comments;
    \item We present the \name concept that uses LLMs to generate tests from LLM-generated comments, and evaluate how many of the generated tests pass. Experiments reveal that our approach has a strong statistical relationship with comment accuracy.
\end{itemize}

The remainder of the paper is organized as follows. We motivate our work in \Cref{sec:motivation}. Our categorization of inaccuracy patterns is given in \Cref{sec:taxonomy}, and \Cref{sec:approach} outlines a technique to detect behavioral inaccuracies. The setup and results of the experiments are given in \Cref{sec:expr_setup} and \Cref{sec:results}, respectively. Finally, \Cref{sec:discussion} discusses caveats and future work, relevant literature is described in \Cref{sec:relwork}, and \Cref{sec:conclusion} concludes.

\section{Motivation}
\label{sec:motivation}

\subsection{Manual Evaluation of Factual Accuracy}

As far as we are aware of, it is difficult to find evaluation of factual accuracy of automatically generated comments. By factual accuracy, we mean that a comment contains no sentences or phrases that describe the behavior or intent of the code falsely; examples are provided with detailed analysis in \Cref{sec:taxonomy}. Despite the importance of evaluation of this nature, supported by the survey results of Aghajani et al.~\cite{aghajani2020docsurvey}, most automatic comment generation techniques opt to use other metrics like BLEU or ROUGE, as mentioned in the previous section~\cite{geng2024llmdoc, mu2023dome, hu2018deep}. Metrics such as BLEU measure the n-gram similarity between a ground-truth answer and generated answer~\cite{papineni2002bleu}. In the context of comment evaluation, one would compare the human-written comment and the automatically generated comment. However, as is soon demonstrated in this section, BLEU has little relation to the factual accuracy of comments generated by LLMs. In this context, the efforts of Mu et al.~\cite{mu2023dome} and Haque et al.~\cite{haque2022semantic} are noteworthy, as they purport to evaluate the accuracy of generated comments via human studies. It is thus unfortunate that their efforts are ultimately difficult to interpret, as both evaluate `accuracy' on an ordinal Likert scale, unlike our understanding of factual accuracy as a black-or-white Boolean concept. We argue that our definition is clearer than a vague evaluation of accuracy on a Likert scale, as it is unclear what it means for a comment generation technique to have an average accuracy score of 3.41~\cite{mu2023dome}, whereas it is clear to say that e.g. 20\% of comments generated by a technique contain a factual error.

An example of an inaccurate comment is one that says `this method triggers an exception when the input is zero' for a method that does not in fact trigger an exception when the input is zero. Such a document objectively and verifiably mischaracterizes the actual behavior of the code, and thus would be misleading and potentially bug-inducing (for example, a developer might add a catch for an exception, expecting to implement different behavior for zero, but that different behavior may never be executed, with different logic being used instead). 

With this in mind, we believe it is important to evaluate automatically generated comments using our definition of factual accuracy. To this end, we sampled 180 public methods from Java projects featured in Defects4J~\cite{Just:2014aa}, and specifically from files that were fixed in each bug in Defects4J. We sampled public methods as they provide the public side of the API, and thus it is likely important to ensure that the documentation for these methods is accurate, as they would impact both developers and others who build upon these projects. Meanwhile, we sampled methods from files that were fixed as such files are more likely to be related to important functionality, relative to files that are not updated. Among the methods in the files, we chose the methods with the longest comments both to ease the labeling process (having a detailed reference helps) and because methods with long comments are of potential importance (developers likely wrote a detailed comment for a reason).

We evaluated LLM-generated comments, as prior work suggests they have the best comment generation performance~\cite{geng2024llmdoc}. While Geng et al.~\cite{geng2024llmdoc} propose a specific prompting technique to further improve LLM performance (in terms of BLEU), we prompted the LLM by simply presenting the method (without comments) and asking it to generate a method-level document, as we wanted to represent the accuracy of the LLM when used in a simple way, instead of analyzing characteristics of outputs from the algorithm of Geng et al. The specific prompt is provided in our supplementary material. Comments were generated using three LLMs: StarCoder~\cite{li2023starcoder}, as an open-source LLM; GPT-3, as one of the most commonly used LLMs~\cite{fan2023large}; and GPT-4~\cite{openai2023gpt4}, to analyze the accuracy of state-of-the-art LLMs, for a total of 540 code comments analyzed. To analyze the comments, the first author went through each LLM-generated comment and compared it with the human comment, the method implementation, and methods that the target method called, to verify the details presented in the comment. Each comment was given two labels: whether any statement within it was inaccurate or not, and whether the judgement was unambiguous (some methods and comments were difficult to judge due to the domain-specific knowledge required to understand them). The second author independently labeled a subset of the comments and discussed the results with the first author; the first author's labels matched the final discussed labels in 87\% of all examined cases.

\begin{figure}
    \centering
    \includegraphics[width=0.9\linewidth]{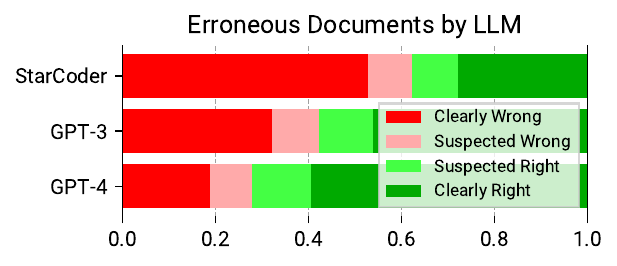}
    \caption{Comment factual accuracy by generating LLM.}
    \label{fig:llm2acc}
\end{figure}

The results of this evaluation are presented in \Cref{fig:llm2acc}. At least half of the comments from StarCoder, a third of the comments from GPT-3, and a fifth of the comments from GPT-4 contained statements that contradicted the written human intent of the method or inaccurately described the behavior of the code. This shows it is difficult to entrust current-generation LLMs with comment generation at scale, as they are prone to inaccurately describing behavior in a significant portion of the generated comments.

\begin{tcolorbox}[boxrule=0pt,frame hidden,sharp corners,enhanced,borderline north={1pt}{0pt}{black},borderline south={1pt}{0pt}{black},boxsep=2pt,left=2pt,right=2pt,top=2.5pt,bottom=2pt]
    A manual evaluation of 540 LLM-generated comments shows they are prone to generating inaccurate comments for Java methods.
\end{tcolorbox}

\subsection{Predicting Factual Accuracy of Comments with Existing Work}

\begin{table}[]
    \centering
    \scalebox{0.9}{ 
    \begin{tabular}{lrrr}
        \toprule
        Technique & Welch's t-test & Point-Biserial Corr.~\cite{lev1949pointbiserial} \\\midrule
        DocChecker~\cite{Dau2024DocChecker} & 0.735 & 0.738 \\
        Deep-JIT~\cite{panthaplackel2021deep} & 0.207 & 0.246 \\
        GPT-3-NoCoT~\cite{li2024mutation} & 0.249 & 0.227 \\
        GPT-3-CoT & 0.168 & 0.156 \\
        BLEU~\cite{papineni2002bleu} & 0.385 & 0.405 \\
        SentenceBERT~\cite{haque2022semantic} & 0.177 & 0.163 \\
        CodeT5~\cite{wang2021codet5} & 0.669 & 0.657 \\
        CodeBERT~\cite{feng2020codebert} & 0.254 & 0.275 \\
        CID~\cite{tanzil2024cid} & 0.747 & 0.748 \\
        \bottomrule
    \end{tabular}}
    \caption{Statistical relationship (p-values) between existing consistency models and factual accuracy of comments.}
    \label{tab:baseline_pvalue_results}
\end{table}

In theory, code-comment consistency detection techniques should be able to uncover a large portion of factual accuracy issues. To test this, we measured whether there was a statistically significant relationship between the model output of nine different baselines, and the factual accuracy of GPT-3 generated comments which we could unambiguously label (141 in total). The baselines are as follows: first, from code-comment consistency detection techniques, we used Deep-JIT~\cite{panthaplackel2021deep}, DocChecker~\cite{Dau2024DocChecker}, and Li and Shin~\cite{li2024mutation}. Deep-JIT is a technique primarily focused on whether a comment should be updated given a program patch, but they also report plain code-comment consistency results in their `post hoc' setting. As they do not provide a trained model, we trained one anew with their public code using their post hoc setting, and confirm the accuracy is similar to that reported within their paper. DocChecker is a recent state-of-the-art technique that builds upon Deep-JIT. Li and Shin suggest asking the LLM itself whether code and comments are inconsistent (marked as GPT-3-NoCoT in \Cref{tab:baseline_pvalue_results}); we additionally add a variant of Li and Shin's technique using Chain-of-Thought (CoT), which is known to improve performance in LLM problem solving~\cite{Wei2022ChainOT}. Second, we evaluate whether similarity metrics can predict accuracy. For the first two, BLEU~\cite{papineni2002bleu} and SentenceBERT~\cite{reimers2019sbert}, we evaluate the correlation between the generated comment accuracy and embedding similarity of the human-written and generated comment. As Haque et al.~\cite{haque2022semantic} find a correlation between SentenceBERT similarity and `subjective' correctness from humans, SentenceBERT allows us to indirectly compare our `factual' correctness labels with subjective correctness. For the latter two, CodeT5~\cite{wang2021codet5} and CodeBERT~\cite{feng2020codebert}, we evaluate the correlation between the comment accuracy and embedding similarity of the target method and generated comment, as these models are trained on source code. Finally, we evaluate the LLM-based inconsistency detection technique CID~\cite{tanzil2024cid}, which was proposed to find inaccurate responses in question-answering tasks. 

To check the significance of the relationships, we use two tests. First, we use the Welch's t-test to evaluate whether a model's output is significantly different for accurate and inaccurate comments, which would be a precondition for distinguishing accurate comments. Second, we use the Point-Biserial correlation coefficient, which is a correlation metric used when one variable is dichotomous (i.e. binary), as is the case in our accurate/inaccurate labels.

The results of this analysis are presented in \Cref{tab:baseline_pvalue_results}. It is notable that there was no statistically significant relationship among the multiple baselines that were tested. This suggests how difficult it is to automatically identify whether an LLM-generated comment contains errors or not - none of the code-comment inconsistency detection techniques, nor any of the LLM self-inspection techniques proposed by the literature detected the problems, nor could existing metrics such as BLEU and SentenceBERT reliably measure accuracy, even with access to the ground truth document.

\begin{tcolorbox}[boxrule=0pt,frame hidden,sharp corners,enhanced,borderline north={1pt}{0pt}{black},borderline south={1pt}{0pt}{black},boxsep=2pt,left=2pt,right=2pt,top=2.5pt,bottom=2pt]
    The output of existing techniques shows little relationship with the factual accuracy of comments.
\end{tcolorbox}

\section{Taxonomy and Examples of Inaccuracy Patterns}
\label{sec:taxonomy}

\begin{figure}
    \centering
    \includegraphics[width=1.0\linewidth]{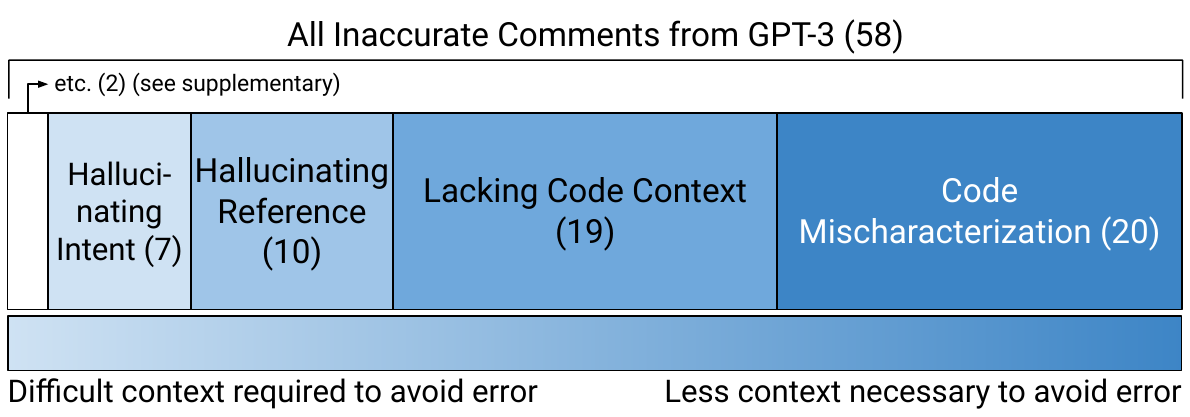}
    \caption{Diagram of error taxonomy for GPT-3 comments.}
    \label{fig:taxonomy_diagram}
\end{figure}

Until now, inaccurate comments have only been described in the abstract. In this section, we detail the specific error patterns that appear in LLM-generated comments by categorizing them, and providing specific examples of erroneous comments to help understanding of this issue. In particular, we analyze the inaccurate comments generated by GPT-3 labeled in the previous section. The results of this analysis are presented in \Cref{fig:taxonomy_diagram}, and are presented in order of increasing severity of error.

The first type of error relates to inaccurately describing the \emph{intent} of the code, in which the intended use description of the method is in conflict with what the developer wrote. Accurately generating comments that describe the intended use is particularly difficult for LLMs, as they cannot access a developer's mind; it is also difficult to automatically verify these comments, outside of asking a developer. An example of this type of error is provided in \Cref{tab:hallucinating_intent_example}. Here, the supported options are inaccurately described - while the LLM-generated comment suggests all options in the SevenZMethod enum are supported, the human comment limits the supported types to a subset of them. While this is clearly an error which could mislead developers, this is not apparent from the code method or the code class (not shown in \Cref{tab:hallucinating_intent_example}). As such, while clearly an error, it is difficult to envision the context that one would need to provide to an LLM to avoid this error.

\begin{table}[]
    \centering
    \scalebox{0.9}{ 
    \begin{tabular}{l|p{7.5cm}}
        \toprule
        Human & Sets the compression method to use for entry contents - the default is LZMA2. Currently only {@link SevenZMethod\#COPY}, {@link SevenZMethod\#LZMA2}, {@link SevenZMethod\#BZIP2} and {@link SevenZMethod\#DEFLATE} are supported. \\\midrule
        LLM & Sets the compression method to be used for compressing the content of a SevenZ file. @param method the compression method to be used. Must be one of the options available in the {@link SevenZMethod} enum.\\
        \bottomrule
    \end{tabular}}
    \caption{Example of the `Hallucinating Intent' error type}
    \label{tab:hallucinating_intent_example}
\end{table}

\begin{table}[]
    \centering
    \scalebox{0.9}{ 
    \begin{tabular}{l|p{7.5cm}}
        \toprule
        LLM &  ... This list contains all the options that have been added using the \{@link \#addOption(Option)\} method ...\\
        \bottomrule
    \end{tabular}}
    \caption{Example of the `Hallucinating Reference' error type}
    \label{tab:hallucinating_reference_example}
\end{table}

The second type of error hallucinates an inaccurate reference to the code, potentially leading developers to look for nonexistent resources. An example of this type of error is provided in \Cref{tab:hallucinating_reference_example}. This is a comment generated for a method named \texttt{getOptions}. For this method, the LLM generated a comment saying that this method will return all options that were added by the \texttt{addOption} method. However, there is no such method in the relevant classes.

\begin{table}[]
    \centering
    \scalebox{0.9}{ 
    \begin{tabular}{l|p{7.6cm}}
        \toprule
        LLM &  ... The type node information includes start and end locations in the source code, as well as whether or not the type node is a simple name or a more complex type expression. ...\\
        \bottomrule
    \end{tabular}}
    \caption{Example of the `Lacking Code Context' error type}
    \label{tab:lack_context_example}
\end{table}

The third type of error relates to the comment inaccurately describing behavior that is caused by methods called by the target method, which are not immediately apparent from the prompt given to the LLM. An example of this type of error is provided in \Cref{tab:lack_context_example}. Here, a \texttt{Node} type object is returned, and the LLM generates an explanation that the returned object includes the start and end locations in the source code. However, if one actually inspects the source code of the Node object, it does not contain any fields that are relevant to start and end locations, and thus the automatically generated description is wrong. It also demonstrates that while it is possible that identifying context for generating comments could improve accuracy, it would not be trivial. In this example, one may need to provide the full class information of the Node object, to cover for the fields, methods, and the behavior of the methods. The methods of other classes are called within the method as well, so one can imagine the amount of additional context needed to (only potentially) mitigate these issues balloon. 

The final type of error is code mis-characterization, in which behavior clearly provided in the code is nonetheless inaccurately described by the LLM. In such cases, because the LLM itself is making an error, it is difficult to do much other than detect and discard these erroneous comments. It also further questions how much help simply providing more context would do in the `Lacking Code Context' type of errors - if the LLM makes mistakes in describing the behavior of a single method, simply providing more context may not be a panacea. An example of this type of error is provided in \Cref{tab:mischaracteriz_example}. In this example, when the shape parameter \texttt{a} is not positive, the return value is \texttt{Double.NaN}, and no exception is thrown. Despite the LLM correctly recognizing the parameter \texttt{a} as the shape parameter, it nonetheless inaccurately describes that a MathException is thrown when the shape parameter is not positive.

\begin{table}[]
    \centering
    \scalebox{0.9}{ 
    \begin{tabular}{l|p{7.9cm}}
        \toprule
        Code &     if (... || (a <= 0.0) || ...) \{ ret = Double.NaN; \} \\
             & ... \\
             & return ret; \\ \midrule
        LLM & ... @param a the shape parameter of the Gamma distribution ...\\
            & @throws MathException if the shape parameter is not positive or if the upper bound of the integral is negative ...\\
        \bottomrule
    \end{tabular}}
    \caption{Example of the `Code Mischaracterization' type}
    \label{tab:mischaracteriz_example}
\end{table}

Noteworthy in this categorization is that while the `Hallucinating Intent' and `Hallucinating Reference' categories generally require external supervision to verify and are thus difficult to automatically detect, the two more common categories, `Lacking Code Context' and `Code Mischaracterization', can often be detected via ``testing'' the documents. For example, for the comment in \Cref{tab:mischaracteriz_example}, one could check if a \texttt{MathException} is triggered when the shape parameter is not positive and thus verify the comment content. This leads to the key question of our work: instead of directly asking LLMs whether a comment is consistent or not, which we found to be ineffective in the previous section, could LLMs be used to generate \emph{tests} to automatically verify LLM-generated comments?

\section{Detecting Incorrect Documents}
\label{sec:approach}

In this section, we suggest a concept we call \emph{\name}, in which the factual accuracy of automatically generated comments is estimated by generating tests from those comments using LLMs and executing them. We first describe the intuition behind our approach, then describe our specific prompting and test execution pipeline to implement this intuition.

\subsection{Intuition}
\label{sec:math_model}
The high-level intuition of our approach is that when prompting an LLM to generate tests, an accurate comment will lead to the LLM generating a higher proportion of passing tests (as the written behavior is accurate), and vice versa; if this assumption holds, we would then be able to distinguish which comments are accurate or not by observing test results. 

To be specific, we provide a mathematical model of our technique, using the terminology of Bayesian inference. Let us define the following: an LLM can generate a document $d$, which may or may not be in the set of all accurate documents $D^+$. To say a document is accurate, we will denote that $d \in D^+$. Next, we denote the probability that an LLM will generate a passing test given a document $d$ as $P(\text{pass}|d)$. We can thus denote the probability that an LLM will generate a passing test given a correct document as $P(\text{pass}|d \in D^+)$.
With this notation, we rewrite our intuition in \Cref{eq:assumption}. Note that we do not assume the LLM generates tests perfectly - we only assume that accurate documents lead to a higher probability of generating passing tests than inaccurate documents.

\begin{equation}
    P(\text{pass}|d \in D^+) > P(\text{pass}|d \notin D^+)
    \label{eq:assumption}
\end{equation}

Usually, we do not know if a document is accurate or not (before manual inspection), but we can automatically know if a test derived from a document passed or failed. Thus, the probability of interest is the probability that a document is accurate, given that a test has passed, $P(d \in D^+ | \text{pass})$. Using Bayes' rule, this can be expressed as

\begin{equation}
    P(d \in D^+ | \text{pass}) = \frac{P(\text{pass}|d \in D^+) \times P(d \in D^+)}{P(\text{pass})}
\end{equation}

\noindent In practice, the odds form of Bayes' rule is easier to use, which is:

\begin{equation}
    o(H|E) = \frac{P(E|H)}{P(E|\neg H)} o(H)
\end{equation}

\noindent where $o(x)=\frac{p(x)}{1-p(x)}$ is the odds function. In the document testing scenario, one would have the following:

\begin{align}
    o(d \in D^+|\text{pass})&=\frac{P(\text{pass}|d \in D^+)}{P(\text{pass}|d \in D^-)}o(d \in D^+) \\
    o(d \in D^+|\text{fail})&=\frac{P(\text{fail}|d \in D^+)}{P(\text{fail}|d \in D^-)}o(d \in D^+) \\
                            &=\frac{1-P(\text{pass}|d \in D^+)}{1-P(\text{pass}|d \in D^-)}o(d \in D^+)
\end{align}

For simplicity, let us set $p_1 = P(\text{pass}|d \in D^+)$ and $p_2 = P(\text{pass}|d \notin D^+)$. The strength of Bayes' rule, and the odds form of it in particular, is that multiple bits of evidence can be combined to a single estimate. For example, say that for a certain document, one test passed and one test failed. Then

\begin{align}
    o(d \in D^+|\text{pass, fail}) &= \frac{p_1}{p_2}o(d \in D^+|\text{fail}) \\
                                   &= \frac{p_1}{p_2}\frac{1-p_1}{1-p_2}o(d \in D^+)
\end{align}

\noindent Generalizing from this, for $n_p$ passing tests and $n_f$ failing tests generated from document $d$, the odds of $d$ being accurate would be

\begin{equation}
    o(d \in D^+|\text{pass=}n_p,\text{fail=}n_f) = (\frac{p_1}{p_2})^{n_p}(\frac{1-p_1}{1-p_2})^{n_f}o(d \in D^+)
\end{equation}

\noindent Taking a log over both sides, and using the fact that the odds and log function are monotonically increasing functions, we can derive an estimator directly proportional to the probability that a document would be accurate, according to Bayesian inference:

\begin{align}
    P(d \in D^+|\text{pass=}n_p,\text{fail=}n_f) &\propto o(d \in D^+|\text{pass=}n_p,\text{fail=}n_f) \\
                                                 &\propto \text{log}(o(d \in D^+|\text{pass=}n_p,\text{fail=}n_f)) \\
                                                 &\propto n_p \text{log}(\frac{p_1}{p_2}) + n_f \text{log}(\frac{1-p_1}{1-p_2}) \\
                                                 &\propto n_p - w n_f  (\because \text{log}(\frac{p_1}{p_2}) > 0)\label{eq:final_estimator}
\end{align}

\noindent where $w=-\text{log}(\frac{p_1}{p_2})/\text{log}(\frac{1-p_1}{1-p_2})>0$, and the prior term $o(d \in D^+)$ is omitted because it is the same over all comments. Simply put, the probability that a document is accurate according to our assumptions is proportional to the number of passing tests subtracted by a weight parameter $w$ times the number of failing tests, as in \Cref{eq:final_estimator}. We call this derived estimator the \emph{correctness estimator}, and use this terminology throughout the paper.

\begin{figure}
    \centering
    \includegraphics[width=\linewidth]{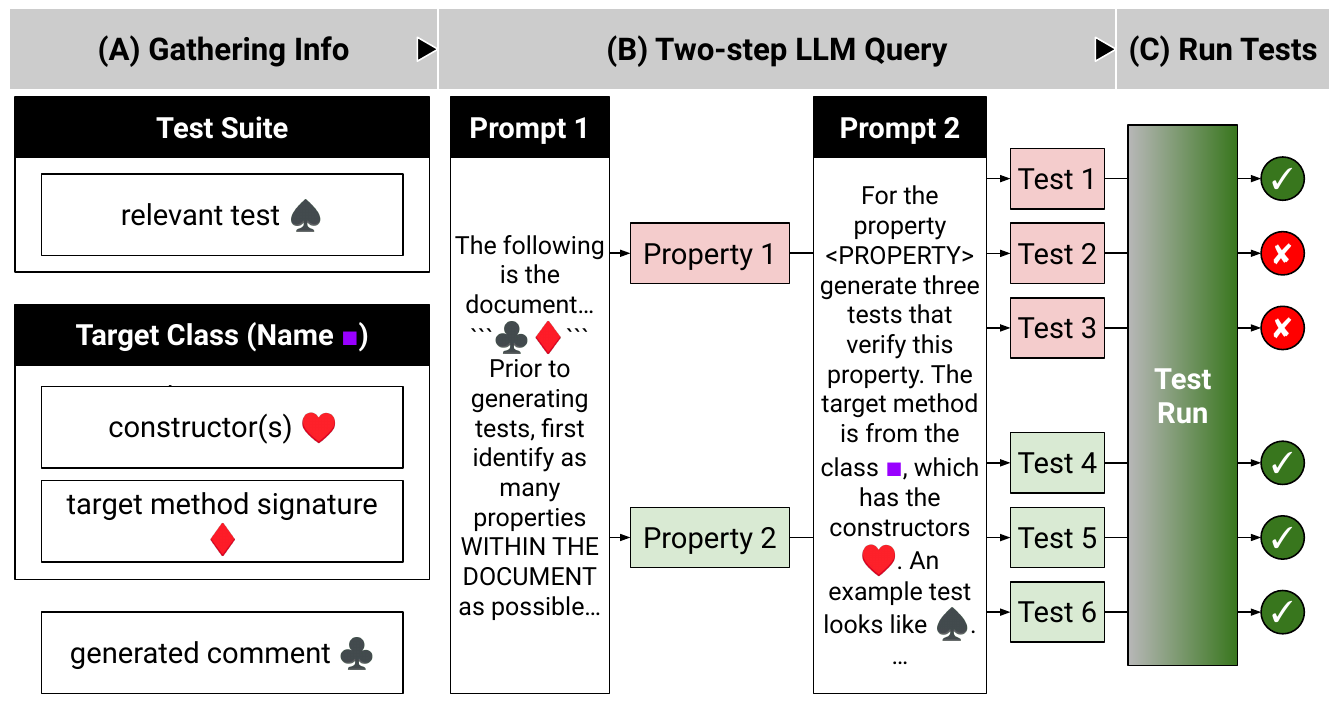}
    \caption{Diagram of document testing pipeline.}
    \label{fig:pipeline_diagram}
\end{figure}

\subsection{Implementation}
An overview of our pipeline, focused on verifying method-level comments, implementing \name is presented in \Cref{fig:pipeline_diagram}. First, along with the generated comment and the signature of the target method, 
other relevant information from the repository is collected to reduce the influence of environment complexity and in the hopes of having the test quality mainly depend on the generated comment (\Cref{fig:pipeline_diagram} (A)). With this information, an LLM is prompted in two stages - first, it is prompted to extract the executable properties given in the comment, then prompted to generate tests that exercise these properties (\Cref{fig:pipeline_diagram} (B)). Finally, the generated tests are injected into the existing test suite and executed to see whether they pass or fail, allowing one to estimate the quality of the comment (\Cref{fig:pipeline_diagram} (C)).

\subsubsection{Information Retrieval from the Repository}
Given the comment we seek to evaluate and its target method, we first extract the signature of the target method. Here, only the signature is provided because the aim is to evaluate the comment; if the code itself is exposed, the LLM may be more influenced by the code than the comment when generating tests, and thus accurate evaluation of the comment itself could become difficult. Meanwhile, the signature, which includes the name and arguments of the method, is necessary to accurately trigger the target method and exercise the functionality in the comment. In addition to this, class-specific information, specifically the name of the class and constructors, is extracted to help with test generation. The reason for this is that without the class information, in many cases the object needed to trigger the target method cannot be constructed, or is inaccurately constructed, causing tests to fail to compile. Similarly, constructors are required to set up the test to exercise the target method.

Finally, we extract example tests related to the target method from an existing test suite. We experiment with two ways of getting relevant tests. The first is retrieving tests from the developer-written test suite. While providing class names and constructors helps, perhaps the best way to help the LLM would be to provide examples of the method or class being used, which it could then modify appropriately.
To identify relevant tests which could help with this process, we employ two heuristics. First, tests are examined to check whether they contain the class name, as tests with the class name likely contain the object initialization process necessary to call the target method. Second, tests are examined to check whether they call a method with the same name and number of arguments as the target method. In such cases, while the method being called in the test may not be the same as the target method due to method overloading in Java, the intuition is that lexically similar methods would likely be used in similar ways, providing a reasonable reference for the LLM. With all tests labeled, we prioritize tests that have both the class name and the associated method call, then tests that have only the class name, then tests that have only the method call. The rationale behind this ordering is that the object initialization process is generally more complex than the act of calling the method.

We acknowledge that test suites are not always available, and also experiment with using tests from the test generation technique EvoSuite~\cite{Fraser:2013vn}. In this case, we set the target class to the class where the target method resides, allow EvoSuite to generate tests, and then use all tests that contain the target name with the appropriate number of arguments as an example. When using EvoSuite-generated tests in the prompt, we replace all strings and integers to dummy values, as we find the LLM often copies the randomly generated strings and integers of EvoSuite, reducing its effectiveness in exercising the comment.

\subsubsection{Two-stage LLM Prompting}
Large Language Models (LLMs), which are statistical models that are trained on large datasets to predict the next token, have shown strong performance in many software engineering tasks, including generating tests from natural language~\cite{kang2023large,xia2023fuzz4all}. Using their capability to follow instructions and perform natural language processing tasks, we `prompt' or instruct them by providing the information gathered in the previous section.
In the first stage of prompting, the generated comment and method signature are given in the prompt, and we request the LLM to extract testable properties from the comment, of the form ``WHEN [condition], THEN the method does [behavior]''. As comments can vary significantly in length and descriptiveness, we leave the LLM to decide how many properties are of interest within the comment. The full prompt is provided in the supplementary material.

For each of these properties, we then provide the target class, class constructor, and the example test(s) in the prompt, and request the LLM to generate three tests that exercise the property. Many properties can be exercised with multiple different inputs, which is why we ask it to generate three separate tests. The hope is that the LLM will thoroughly exercise the generated properties, to improve its capability of distinguishing correct and incorrect properties.

We perform prompting in two stages to improve both performance and usability for practitioners. As we demonstrate in \Cref{sec:rq2_results}, two-stage prompting leads to better performance; qualitatively, we found that without two stage prompting, LLMs would often generate tests unrelated to the actual content of the comment (although we do not claim to have eliminated this phenomenon; see \Cref{sec:rq4_results}). On the other hand, explicitly identifying properties helps developers understand the intent of generated tests and their relationship to the comment, improving usability on their end.

\subsubsection{Test Execution}
To execute the tests, we run a pipeline that performs the following steps, inspired by prior work on bug reproduction~\cite{kang2023large}. First, the token-wise similarity between the generated test and all test files is calculated, and the test is injected into the test file with the greatest similarity. In this process, \texttt{import} statements are added to the highlighted test file if necessary. With this correction done, the test is compiled and executed using the Defects4J command line interface. Using this pipeline, we record whether tests compile, execute, and whether they pass. 
The final score given to each comment is the correctness estimator from \Cref{eq:final_estimator}. For the purposes of our experiment, tests that do not compile are ignored, as they cannot be judged as either passing or failing.

\section{Experimental Setup}
\label{sec:expr_setup}
We describe our experimental setup.

\subsection{Research Questions}
\noindent \textbf{RQ1: How well can LLM-generated tests identify inaccurate comments?}
In this research question, we seek to understand whether LLM-generated tests can help identify inaccurate comments. To do so, we first confirm whether our starting assumption that the test pass rate would be higher for accurate documents is correct, then evaluate the performance of the correctness estimator of \Cref{eq:final_estimator} via the ROC-AUC and AP metric.

\noindent \textbf{RQ2: How much does the helper information provided in the prompt improve performance?}
This research question confirms whether the components included in the prompt as additional information indeed help in distinguishing incorrect documents. Specifically, does adding the class name, constructors, and example tests actually help the LLM generate more executable tests and distinguish inaccurate comments? We compare the number of executable tests (i.e., tests that compile) and the ROC-AUC value of different settings of our implementation.

\noindent \textbf{RQ3: How does the $w$ parameter influence performance?} The comment correctness estimator (\Cref{eq:final_estimator}) has a parameter $w$, which determines whether to place a greater weight on passing or failing tests when estimating the likelihood that a comment is incorrect. We plot how well accurate comments can be distinguished by varying $w$ over experiments.

\noindent \textbf{RQ4: How could the generated tests help developers assess automatically generated comments?} In this research question, we seek to provide a qualitative analysis of the successes and failures of our approach, highlighting its potential and limitations.

\subsection{Evaluation Details}
For our study, we focused on whether GPT-3 (\texttt{gpt-3.5-turbo-0125}) could be used to evaluate comments that it generated itself. Thus, the 141 unambiguously labeled comments and methods from \Cref{sec:motivation} were used as subjects for our experiment. We did so because we wanted to provide a proof of concept that documentation testing could work with a single LLM, and that it would not require a complex setup. Unless specified otherwise, we repeated experiments five times to verify that our approach works consistently.

For RQ1, we report the average test passing rate, to verify our starting assumption made in \Cref{sec:motivation}, and also in assessing the effectiveness of our prompting technique. We also use the correctness estimator from \Cref{eq:final_estimator} in RQ1 to RQ3. We evaluate metrics on the basis of the p-values for the Welch's t-test and Point-biserial correlation, as well as the ROC-AUC and the average precision (AP) value of our approach; the ROC-AUC value is commonly used to evaluate binary classifiers as a general measure of predictiveness~\cite{fawcett2006rocauc}, while AP is suggested when the classes are unbalanced~\cite{yuan2015threshold}. ROC-AUC is 0.5 for a random classifier, while AP depends; for both, higher values are better. For RQ1, we additionally report the proportion of accurate comments within a given estimator score range. For RQ2, we additionally report the percentage of comments for which at least one executable test was generated, as an indication of how effective the prompting techniques were in providing useful information to generate tests. $w=100$ was used in RQ1 and RQ2.

\section{Results}
\label{sec:results}
This section outlines the results of experiments performed.

\subsection{RQ1: Predicting Inaccuracies}
\label{sec:rq1_results}
\begin{figure}
    \centering
    \begin{subfigure}[t]{0.49\linewidth}
        \centering
        \includegraphics[width=\linewidth]{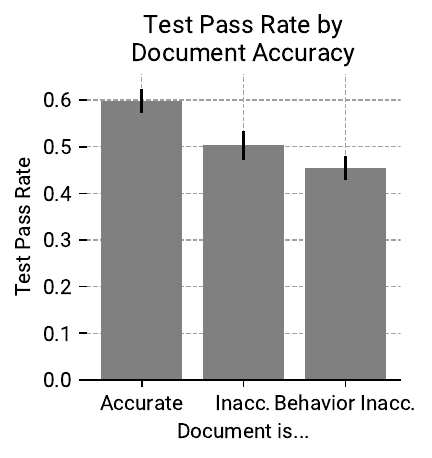}
        \caption{Pass rate by accuracy, with 95\% confidence intervals.}
        \label{fig:docacc2testpass}
    \end{subfigure}
    \begin{subfigure}[t]{0.49\linewidth}
        \centering
        \includegraphics[width=\linewidth]{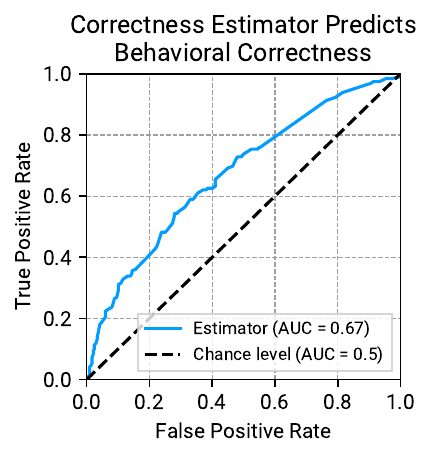}
        \caption{ROC graph of correctness estimator with actual correctness.}
        \label{fig:correctness_estimator_roc}
    \end{subfigure}
    \caption{Relationship between comment accuracy and suggested indicators.}
\end{figure}

\begin{figure}
    \centering
    \includegraphics[width=0.95\linewidth]{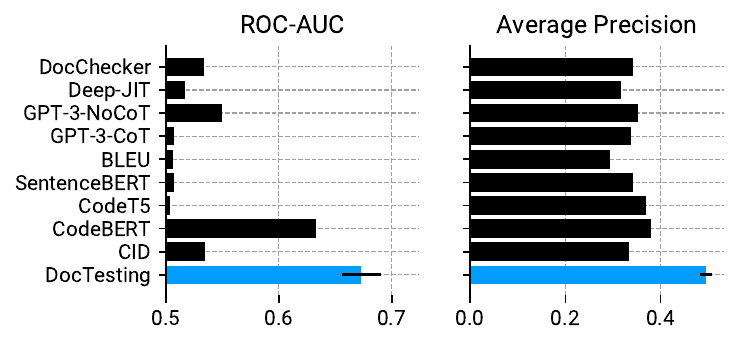}
    \caption{ROC-AUC and AP values compared with baselines. For our approach (blue), we present the mean value from five runs, along with its 95\% confidence interval.}
    \label{fig:rocauc_comparison}
\end{figure}

\begin{figure}
    \centering
    \includegraphics[width=0.75\linewidth]{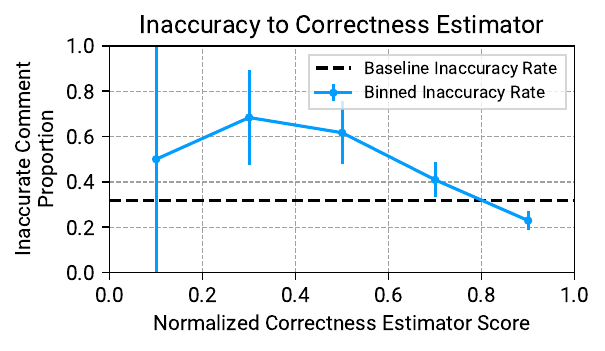}
    \caption{Comment accuracy by normalized estimator value bin, shown with the 95\% confidence interval of each bin.}
    \label{fig:binning_results}
\end{figure}

This research question relates to whether generating and executing tests from comments helps distinguish accurate and inaccurate comments. We first verify our assumption that the test pass rate would be higher for factually accurate comments (\Cref{eq:assumption}). \Cref{fig:docacc2testpass} shows the test pass rate by document type, averaged over the five runs that we did. As presented in the graph, there is a substantial and statistically significant ($p=0.002$) difference in pass rate between accurate and inaccurate documents. As mentioned in \Cref{sec:taxonomy}, however, \name is a concept most suited to revealing inaccurate descriptions of program behavior. Given that, one would expect comments that inaccurately describe the behavior of the code (those labeled lacking context or code mischaracterization in \Cref{sec:motivation}) should have a greater difference. Indeed, this is what we observe - the test pass rate for comments that inaccurately describe the behavior of the code is even lower, and the difference is more significant ($p<10^{-4}$). As this is the intended use case, we exclude the comments that are inaccurate for other reasons in the remainder of our analysis.

Although there is a meaningful difference in test pass rate between accurate and inaccurate comments, recall we derived the correctness estimator in \Cref{sec:math_model}, which would theoretically have a direct correspondence with the likelihood of a document being accurate according to Bayesian inference, unlike test pass rate. In \Cref{fig:correctness_estimator_roc}, we plot the ROC (receiver operating characteristic) curve between our correctness estimator and actual comment correctness, showing a robust relationship between the two. The statistical relationship between comment accuracy and our correctness estimator is even stronger than in the test pass rate case - the point-biserial correlation between the two is extremely unlikely to be due to chance ($p<10^{-11}$), and the Welch's t-test also yields a significant difference for estimator value between correct and incorrect comments ($p<10^{-9}$). We emphasize that no other baseline had a statistical relationship with comment factual accuracy, whereas our estimator shows a strong relationship.

In \Cref{fig:rocauc_comparison}, we compare the ROC-AUC and AP values of different predictors with the values achieved by \name. Document testing does the best in distinguishing correct comments from inaccurate ones, as no other method comes within its 95\% error margins. Along with the significant statistical relationships that we found between the correctness estimator and actual correctness, this makes a strong case that testing comments via LLMs is a viable way of assessing the factual accuracy of documents. Among the baselines, CodeBERT is worthy of further analysis, as it seemingly has a high ROC-AUC value. We have two reasons to believe these results are spurious. First, it does not have a notable AP value. More importantly, while we could not repeat this experiment for CodeBERT as its output is deterministic, we ran CodeBERT on the StarCoder-generated comments from \Cref{sec:motivation} and found a correlation in the \emph{opposite} direction; its ROC-AUC value was 0.35. In contrast, our approach on the same StarCoder-generated comments achieved an ROC-AUC of 0.63, and thus behaved as expected.

We also plot the conditional probability that a comment will be inaccurate based on the correctness estimator value, normalized to the [0, 1] range, in \Cref{fig:binning_results}. With the exception of the $[0, 0.2)$ bin, where there were too few samples for a reasonable confidence interval to be constructed, there is a trend in the expected direction: as the correctness estimator deems a comment more likely to be accurate, the relative proportion of inaccurate comments monotonically decreases, providing an opportunity to filter out inaccurate comments. For example, rejecting comments with a normalized correctness estimator score lower than 0.8 would remove 46\% of the inaccurate comments, while retaining 72\% of the correct comments.

\begin{tcolorbox}[boxrule=0pt,frame hidden,sharp corners,enhanced,borderline north={1pt}{0pt}{black},borderline south={1pt}{0pt}{black},boxsep=2pt,left=2pt,right=2pt,top=2.5pt,bottom=2pt]
    \textbf{Answer to RQ1:} The assumption of document testing, that correct comments will lead to more passing tests, is verified ($p<0.01$); the comment correctness estimator shows a strong statistical relationship to comment correctness as well ($p<10^{-9}$).
\end{tcolorbox}

\subsection{RQ2: Prompt Ablation}
\label{sec:rq2_results}

\begin{figure}
    \centering
    \includegraphics[width=\linewidth]{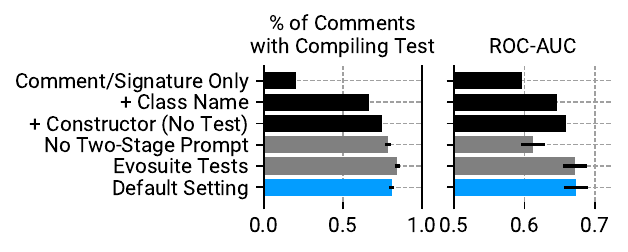}
    
    \caption{Prompt ablation results with 95\% CIs. The top three experiments were performed once, so they have no CIs.}
    \label{fig:ablation_all}
\end{figure}

This research question seeks to verify whether the helper information that we provided in the prompt, along with the document, was indeed useful in generating more executable tests and in distinguishing inaccurate comments. First,  the left graph of \Cref{fig:ablation_all} shows the proportion of comments for which at least a single executable test was generated. As it shows, each element helps improve the number of comments meaningfully processed, enhancing the proportion of situations in which \name could be applied. Meanwhile, the right graph of \Cref{fig:ablation_all} shows how well each ablated case can distinguish incorrect comments using the ROC-AUC metric. Our default setting also performs best here, showing that each component contributes to a high performance. Meanwhile, our experiments with using EvoSuite in lieu of human tests shows that using EvoSuite-generated tests yields comparable performance.

\begin{tcolorbox}[boxrule=0pt,frame hidden,sharp corners,enhanced,borderline north={1pt}{0pt}{black},borderline south={1pt}{0pt}{black},boxsep=2pt,left=2pt,right=2pt,top=2.5pt,bottom=2pt]
    \textbf{Answer to RQ2:} Each part of the helper information aids in generating more executable tests, and thus furthers the ability of \name to distinguish correct and incorrect comments.
\end{tcolorbox}

\subsection{RQ3: Influence of $w$}
\label{sec:rq3_results}

\begin{figure}
    \centering
    \includegraphics[width=0.75\linewidth]{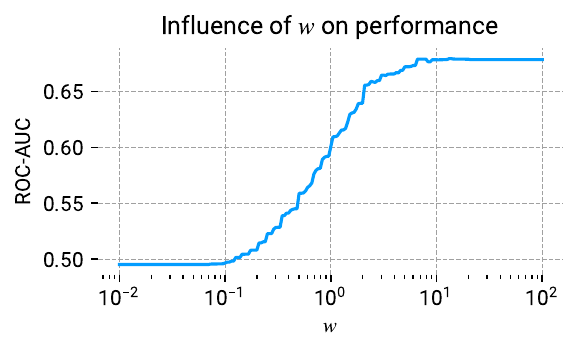}
    \caption{The influence of the parameter $w$ on accuracy prediction performance.}
    \label{fig:weight2perf}
\end{figure}

In this research question, the influence of the parameter $w$, which determines how much weight will be placed on failing tests relative to passing tests in the correctness estimator, is investigated. We vary $w$ exponentially using the schedule $w_i = 100^{\frac{i}{100}-1}$ where $i \in \{0, 1, ..., 200\}$, and evaluate the discriminative power of documentation testing using the ROC-AUC value. This splits the space between $\frac{1}{100}$ and $100$ evenly on a logarithmic scale. The results of this experiment are presented in \Cref{fig:weight2perf}. Interestingly, we find the best performance (ROC-AUC of 0.67) when $w$ is high ($w>10$), meaning that the number of the failing tests is the most predictive of inaccurate comments, while the number of passing tests is best used as a tiebreaker. This is likely explained by our comment labeling process. As mentioned in \Cref{sec:motivation}, even if a comment is mostly correct, if it contains even one description that is factually inaccurate, the comment was rated as inaccurate. Due to this, it is possible for a comment to lead to many passing tests, and still be inaccurate. In other words, even a small proportion of failures may be indicative of a problem in the comment, as presented next.

\begin{tcolorbox}[boxrule=0pt,frame hidden,sharp corners,enhanced,borderline north={1pt}{0pt}{black},borderline south={1pt}{0pt}{black},boxsep=2pt,left=2pt,right=2pt,top=2.5pt,bottom=2pt]
    \textbf{Answer to RQ3:} Performance is best when $w>10$, suggesting the number of failing tests is more important in predicting comment inaccuracies than the number of passing tests.
\end{tcolorbox}

\subsection{RQ4: Qualitative Study}
\label{sec:rq4_results}

\begin{figure}[h]
\centering
\includegraphics[width=0.45\textwidth]{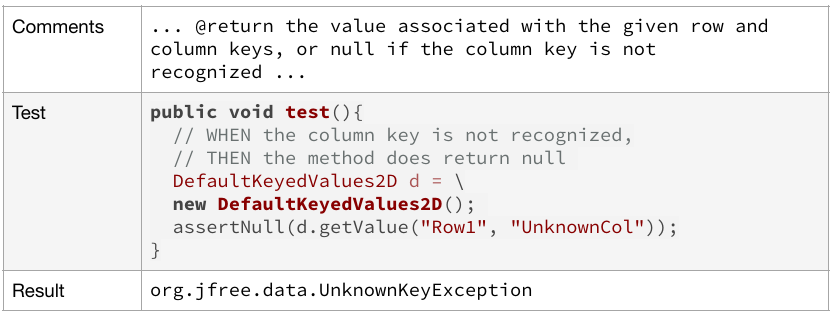}
\caption{A successfully verified LLM-generated comment.}
\label{tab:best_case_example}
\end{figure}

Finally, we qualitatively analyze best-case and worst-case results for our implementation of \name, which shed light on the strengths and weaknesses of our technique. We first present an example that showcases the potential benefits of testing comments via the pipeline we have in \Cref{tab:best_case_example}. In this case, the comment is mostly correct, but inaccurately describes the return behavior, saying that null is returned if the column key is not recognized. Taking cue from this statement, our pipeline generated the test in the `Test' row, which seeks to verify that when a column key is not recognized, the method returns null (note the use of the ``UnknownCol'' string). However, this test fails, because this is an inaccurate description of the behavior - when the column key is not recognized, an UnknownKeyException is triggered instead. In this case, the developer would be able to look at the test, check which specific property of the comment is wrong, and confirm that the description is inaccurate by looking at the test and the execution result. This aspect distinguishes our approach from other code-comment consistency techniques; in many techniques, even when they work, one would only know that there is a high likelihood of the comment being inaccurate, without knowing what in particular is problematic. On the other hand, in our ideal, a developer would not only know whether a comment is wrong, but what about it is wrong, and verify the results via automatically generated tests, ultimately gaining \emph{trust} in the automated process.

\begin{figure}[ht]
\centering
\includegraphics[width=0.45\textwidth]{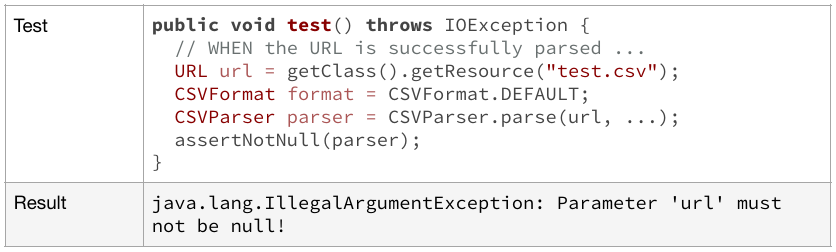}
\caption{A test that fails due to the execution environment.}
\label{tab:worst_case_example_environmental}
\end{figure}

However, we acknowledge that more effort must be put in to reliably meet that ideal. Generally speaking, \name works poorly when the assumption that correct comments will lead to a higher proportion of passing tests breaks down. While we would like to restate that overall this assumption tends to hold as demonstrated in \Cref{sec:rq1_results}, we present two examples where correct comments lead to a high proportion of failing tests, the first of which is provided in \Cref{tab:worst_case_example_environmental}. Here, the test tries to read from a CSV file to construct a URL object, but fails as there is no such file. Thus, \texttt{url} is set to null. However, because a non-null URL is needed to test most properties of the comment, all tests with the exception of those that deal with null \texttt{url} values fail. This points to the need of better test execution environment support to enhance the reliability of results, an issue that can perhaps be improved with better tooling.

\begin{figure}[ht]
\centering
\includegraphics[width=0.45\textwidth]{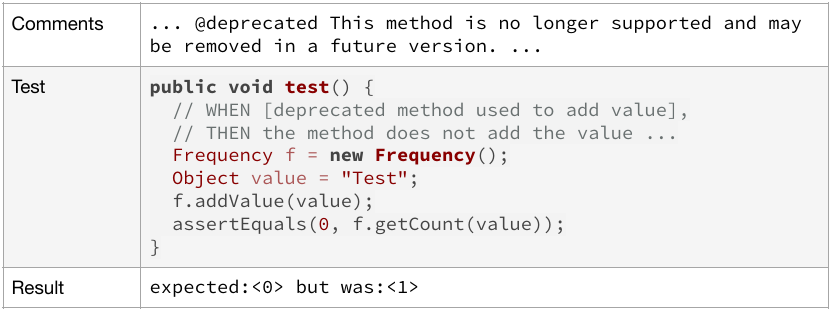}
\caption{A generated test that hallucinates a property.}
\label{tab:worst_case_example_testhallucination}
\end{figure}

A greater concern is that sometimes LLMs will hallucinate properties that are not in the documents. An example is provided in \Cref{tab:worst_case_example_testhallucination}. Here, although the original comment says nothing about program behavior in the @deprecated tag, during the test generation process the LLM made up the property that if the deprecated method is used, it will result in no effect. Thus, the test fails on a correct document, as it is testing a property that has little to do with the original document. While this is a concerning issue, it is one that we expected, as LLMs are known to hallucinate. Indeed, we do not rely on LLMs to be flawless generators of tests; our assumptions only state that correct documents will have a higher chance to lead to passing tests. We could also reduce the occurrence of hallucination in test generation via our two-step prompting process, suggesting that further improvement is possible.

\begin{tcolorbox}[boxrule=0pt,frame hidden,sharp corners,enhanced,borderline north={1pt}{0pt}{black},borderline south={1pt}{0pt}{black},boxsep=2pt,left=2pt,right=2pt,top=2.5pt,bottom=2pt]
    \textbf{Answer to RQ4:} In the best cases, \name can detect inaccurate comments, highlight specifically which part of the comment is wrong, and prove the discrepancy via concrete tests. However, our current implementation also has limitations.
\end{tcolorbox}

\section{Discussion}
\label{sec:discussion}
This section discusses the limitations and implications of our work.

\subsection{Threats to Validity}
\emph{Internal threats} are challenges to the conclusions made within the manuscript. LLM outputs in our work are random and thus key performance metrics such as ROC-AUC can differ between runs. To provide a balanced view, we run our algorithm five times and report the average ROC-AUC value. As the training data of the GPT family of LLMs is unknown, there is also the risk that the LLM had learned the subject code. However, this did not translate into the LLM correctly predicting which comments were accurate on its own (see \Cref{fig:rocauc_comparison}); only with \name could we find a reliable predictor of correctness.

\emph{External threats} are challenges to the generalizability of the reported findings. The abstract mathematical model of verifying documents presented in \Cref{sec:math_model} is agnostic to the underlying programming language, language model, and testing framework. However, the experimental results we present were done on Java code from the widely-used Defects4J benchmark; further research is required to tell whether these principles would work for other languages and projects.

\subsection{Future Work}
There are two levels of generalization that this work could expand into. First, in this paper, the main focus is on detecting inaccurate method-level comments generated by LLMs. 
However, the concept of testing a comment by generating tests from an LLM is not necessarily restricted to LLM-generated documents. After all, in the simplified model presented in \Cref{sec:math_model}, there is no direct notion of `hallucination'; there is only the notion of whether a description is accurate or not. As such, it is feasible that as long as a technique can faithfully translate written documentation into tests, one could use a similar testing-from-comments technique to estimate the quality of human-written documents. Again assuming high fidelity between the comment and generated tests, \name could help developers better understand and debug any differences between the comment and actual behavior, as studies have revealed that developers can perform debugging better with a working example~\cite{Beller2018DebuggingDichotomy}. To make this vision a reality, one would need to increase the comment-to-test fidelity substantially to reduce the number of false positives and increase developer adoption~\cite{arcuri2018experience}. Looking even further, the concept may not even be restricted to comments - perhaps documents such as installation instructions or API references could be verified, using the same principles introduced in this paper.

\section{Related Work}
\label{sec:relwork}
This section introduces relevant research to our efforts.

\subsection{Code-Comment Consistency Detection}
Tan et al.~\cite{tan2007icomment} introduce iComment, a technique that combines simple natural language processing (NLP) and static analysis tools to identify inconsistencies between lock behavior described in the comments and actual implementation. Since then, different methodologies have since been proposed to identify code-comment inconsistencies. One recently popular branch of such approaches are just-in-time comment inconsistency detection techniques~\cite{panthaplackel2021deep}, where a comment is compared with a code patch, and whether the comment should be updated is predicted; in such techniques, deep learning is commonly used~\cite{xu2023data}. While this application is slightly different from our setting in that a code change is required, we run experiments and compare against two state-of-the-art code comment inconsistency detection techniques in our work~\cite{Dau2024DocChecker}. 

While there are many different techniques to detect code-comment consistency, the work of Tan et al.~\cite{Tan2012tcomment} is of particular relevance, which introduces the technique @tComment to extract invariants from method-level comments related to null inputs, and checks if those invariants are violated during execution using the random test generation tool Randoop~\cite{Pacheco2007randoop}. 
While closely related to our hallucination detection approach, noteworthy is the reliance the technique has on the random test generation tool Randoop, which is agnostic to which inputs might trigger special behavior. In this work, we use LLMs to directly trigger and test the behavior described in (potentially inaccurate) LLM-generated comments to confirm their veracity. Additionally, we tried to compare against their work as well, but had difficulty running their tool, which was last updated eight years ago. 

\subsection{Documentation Generation}
Motivated by the need for comments in understanding code, attempts to automatically generate comments and documents have been made as well. Early work includes Buse and Weimer~\cite{buse2010automatically}, which introduced the DeltaDoc technique to generate a document that would describe what change was made in the code. With the advent of deep learning, comment generation was commonly formulated as a translation task (from code to comments), and also increasingly used NLP techniques to generate comments. A good example of this Hu et al.~\cite{hu2018deep}, who used a seq2seq neural architecture to generate comments given Java code. This formulation is still used by recent work - for example, Mu et al.~\cite{mu2023dome} similarly use a translation model, augmented with developer intent such as the intention of the code, and the recent work of Geng et al.~\cite{geng2024llmdoc} use LLMs to achieve the same goal and show LLMs achieve state-of-the-art performance. As mentioned earlier, these techniques either do not evaluate the factual accuracy of their generated documents, or do so in an unsatisfactory way. To overcome this limitation, we manually evaluate the factual accuracy of 540 comments automatically generated by an LLM. It is critical to keep in mind that existing work did not claim that their high BLEU score meant that the comments could be trusted. However, one aim of this work is to clearly report that even automatically generated comments with high BLEU score can be misleading.

\subsection{Hallucination in LLMs}

LLMs are prone to `making up' facts when generating a response, as prominently explored by Shuster et al.~\cite{shuster2021hallucination}. The risks of unquestioningly using the responses of these models has since been clear, and many techniques have been proposed to mitigate these risks. Shuster et al.~\cite{shuster2021hallucination} identified that retrieval augmentation could reduce LLM hallucination in their work. Meanwhile, Wang et al.~\cite{wang2023selfconsistency} suggest that when querying an LLM multiple times with the same query, the number of equivalent responses is correlated with the probability of the LLM to correctly respond, another mechanism that could be used to estimate the likelihood of hallucination. In contrast to these techniques, \name relies on the reliability that code execution can provide. From software engineering literature, CID~\cite{tanzil2024cid} has been proposed to identify inaccurate answers from conversations regarding software library recommendations. Unfortunately, their approach does not seem to generalize to detecting inaccuracies in comments, as we find in \Cref{tab:baseline_pvalue_results}. The recent work of Virk et al.~\cite{virk2024enhancing} argue that rescaled mean token probabilities from LLMs are predictive of human-rated similarity (which Haque et al.~\cite{haque2022semantic} find is correlated to SentenceBERT similarity), but as shown in \Cref{sec:motivation}, SentenceBERT similarity and thus likely human-rated similarity may not correlate with factual accuracy.

\section{Conclusion}
\label{sec:conclusion}
Ensuring the factual accuracy of automatically generated comments is of critical importance for widespread developer adoption of comment generation tools. In this regard, this paper makes three contributions. First, our manual evaluation of 540 generated comments shows that even the best-performing comment generation techniques, LLMs, are prone to generating a substantial portion of factually incorrect comments, demonstrating the practical importance of the problem. Second, we report our experimental results which show little to no relationship between the outputs of existing techniques and the accuracy of generated comments, underscoring the difficulty of the problem. Third, we propose the concept of \name to use LLMs to generate tests from comments to estimate the comment's likelihood of correctness by observing the test execution results, and implement a pipeline to realize the idea. Experiments reveal that there is a robust statistical relationship between the test output results and comment accuracy, validating our approach. In this paper, we focused on LLM-generated method-level comments as a target, but we believe \name has the potential to be expanded to other types of documents. We hope to explore the concept further in future work.

\bibliographystyle{ACM-Reference-Format}
\bibliography{sample-base}

\end{document}